\documentclass[reprint,prx,showpacs]{revtex4-1}
\usepackage{bm}
\usepackage{graphicx}
\usepackage{amsmath,amsfonts,amssymb}
\usepackage{hyperref}

\begin{document}
\title{Manipulation of Dirac Cones in Mechanical Graphene}

\date{\today}
\author{Toshikaze Kariyado}\email{kariyado@rhodia.ph.tsukuba.ac.jp}
\author{Yasuhiro Hatsugai}\email{hatsugai@rhodia.ph.tsukuba.ac.jp}
\affiliation{Division of Physics, Faculty of Pure and Applied Sciences,
University of Tsukuba, Tsukuba, Ibaraki 305-8571, Japan}
%63.22.Rc: phonons in graphene
%03.65.Vf: topological phases, Berry phase
%45.20.D-: Newtonian Mechanics

\begin{abstract}
 Mechanical graphene, which is a spring-mass model with the honeycomb
 structure, is investigated. The vibration spectrum is
 dramatically changed by controlling only one parameter, spring tension
 at equilibrium. In the spectrum, there always exist Dirac cones
 at K- and K'-points. As the tension is modified, extra Dirac cones are
 created and annihilated in pairs.
 When the time reversal symmetry is broken by uniform rotation of the system,
 creation and annihilation of the Dirac cones result in 
 a jump of the appropriately defined Chern number.
 Then, a flip of the propagation direction of the
 chiral edge modes takes place, which gives an experimental way to
 detect the
 topological transition. This is a bulk-edge correspondence of the
 classical system. We also demonstrate the other
 important concept, symmetry protection of the topological states, is at
 work in the classical
 system. For the time reversal invariant case, the topological edge
 modes exist for the fixed boundary condition but not for the free
 boundary condition. This
 contrast originates from the symmetry breaking at the free boundary.
\end{abstract}

\maketitle

\section{Introduction}
Graphene \cite{Novoselov:2005fk}, a two-dimensional crystal with a
honeycomb array of carbon
atoms, has been one of the most hot topics in condensed matter physics
in this
decade. The most prominent feature of graphene is existence of massless
Dirac fermions, or Dirac cones at the Fermi energy. In general, not
limited to graphene, Dirac cones allow
a lot of properties distinct from usual metals or semiconductors, such as
unique responses to electronic and magnetic field
\cite{Zhang:2005aa,Novoselov:2005fk}, or characteristic
edge states \cite{JPSJ.65.1920,PhysRevLett.89.077002}. One interesting
subject of study is a manipulation of Dirac cones. For instance, 
shifting the Dirac cones in momentum space induces
a gauge field without breaking the time reversal symmetry \cite{Guinea:2010aa}. 
By inducing a mass of Dirac cones (gap opening), we have a chance to
observe topologically nontrivial phases. For instance, by appropriately
breaking the time reversal symmetry, graphene can go into a quantum Hall
state
without external magnetic field \cite{PhysRevLett.61.2015}. Or, if the spin
degrees of freedom and the spin--orbit coupling are explicitly taken
account of, graphene becomes a time reversal invariant $Z_2$ 
topological insulator \cite{PhysRevLett.95.146802}. 

One direction of recent developments in Dirac and topological systems 
is exporting the concept to systems other than conventional solid
materials. For instance, manipulation of Dirac cones
is experimentally realized in an optical lattice cold atoms
\cite{Tarruell:2012fk}. 
The other example is electromagnetic field in a photonic crystal, which
is governed by the classical Maxwell equation.
As for triangular and honeycomb lattices, the Dirac cones in photonic
band are demonstrated
\cite{PhysRevB.44.8565,PhysRevB.52.R2217,PhysRevA.75.063813,PhysRevB.80.155103}.
Furthermore, topologically nontrivial phase is achieved in
photonic crystals or a coupled resonator system
\cite{PhysRevLett.100.013905,PhysRevLett.100.013904,HafeziM.:2013aa}. Very recently,
other classical systems, i.e.,
mechanical systems obeying the Newton's equation of motion are also
discussed in the context of Dirac cones or
topological edge states
\cite{PhysRevLett.103.248101,PhysRevE.83.021913,Chen09092014,Kane:2014aa,Po:2014,Coriolis:2014,nash:2015,wang:2015,huber:2015}. One
great advantage of these kinds of
artificial systems is their controllability, which enables us
to access parameter region unreachable in solids.
We can easily realize phenomena that are difficult in solids such as
merging of Dirac cones \cite{springerlink:10.1140/epjb/e2009-00383-0}. 

In this paper, a honeycomb spring-mass model \cite{0143-0807-25-6-004},
we dubbed it as mechanical
graphene, is investigated as a typical mechanical system in which Dirac
physics plays a role. We propose a simple and feasible way, just
stretching the system uniformly and isotropically, to manipulate Dirac
cones. The uniform stretch modifies tension of springs at equilibrium,
which controls the relation between the transverse and longitudinal
waves. Then, the frequency spectrum varies as a function of the
tension. As the tension grows from zero, the frequency band
structure first looks like the one in $p$-orbital honeycomb
optical lattice \cite{PhysRevA.83.023615} (phase I), then becomes like
the one in
bilayer graphene with the trigonal warping \cite{PhysRevLett.96.086805}
(phase II). Finally the spectrum is given by that of
graphene (phase III), but with extra degeneracy. Merging
of the Dirac cones is observed
at the two critical points.
This topological transition associated with
 the merging of the Dirac cones is
detected as a jump in the appropriately defined Chern number
\cite{PhysRevLett.49.405,Coriolis:2014}, when the time
reversal symmetry is broken by uniform rotation of the system. The
change in the Chern number is also observed by the propagation direction
of the ``chiral edge modes'' \cite{PhysRevB.48.11851,PhysRevLett.71.3697}.
Not only proposing a practical method for Dirac cone manipulation, we
also demonstrate the state-of-art idea of topological phenomena,
symmetry protection, is at work in this system by focusing on edge
states and boundary condition. Without the time reversal symmetry
breaking, in-gap edge states appear for the fixed boundary condition,
while they do not appear for the free boundary condition. In the
topological view point, symmetry breaking at the edge is behind this
contrast. 

\section{Model and Methods}
\begin{figure}[tbp]
 \begin{center}
  \includegraphics{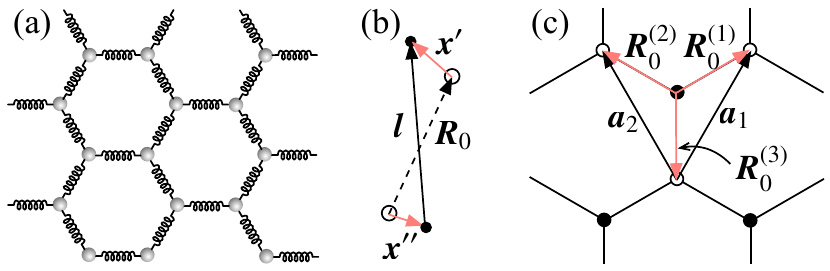}
  \caption{(a) Schematic picture of the mechanical graphene. (b)
  Definitions of $\bm{R}_0$, $\bm{x}'$, and $\bm{x}''$. (c) The unit
  vectors $\bm{a}_1$ and $\bm{a}_2$, and the vectors connecting the
  nearest neighbor sites, $\bm{R}_0^{(i)}$ ($i=1,2,3$). }\label{fig1}
 \end{center}
\end{figure}
The model treated in this paper is composed of mass points, springs, and
additional potential energy sources, where the mass points are aligned
in the two-dimensional honeycomb lattice. [See Fig.~\ref{fig1}(a).]
Motion of the mass points is also
restricted in the two dimensional space, i.e., the out of plane motion
is prohibited. Parameters characterizing this model are mass of the mass
points $m$, a spring constant $\kappa$, distance between the nearest
neighbor mass points $R_0$, and natural length of the springs $l_0$. In
addition, there
are some parameters to characterize additional potential energy other
than the one from springs connecting the mass points. The mass $m$ is
fixed to 1 for simplicity
from now on. $\eta\equiv R_0/l_0$ is not necessarily unity, but when it
deviates from unity, each string gives
finite force even in equilibrium, which means that we should make the 
system to avoid shrinkage as a whole. For instance, we need to support
the system by appropriate choice of the boundary condition. Or, the
potential source other than the springs need to be added. As far as the
shrinkage is prevented, the mass points are stable at the equilibrium
position since the forces from three springs connected to one mass point
cancel out. Dynamical variables for this model is written as
$\bm{x}_{\bm{R}a}={}^t(x_{\bm{R}a},y_{\bm{R}a})$,  where $\bm{R}$
denotes a lattice point, and $a$ is a sublattice index. As we will see
below, the model has commonality with the nearest neighbor tight-binding
model of real graphene, but the major difference is that the mass point
allows to oscillate in two directions, $x$ and $y$. There is also the
sublattice degrees of freedom, which means that the frequency spectrum
for this model has four bands.

For this model, elastic energy of the spring for a given pattern of mass
point configuration has a special importance. We first see it for a pair of
mass points, introducing parameters $\bm{R}_0$, $\bm{x}'$, and $\bm{x}''$
as shown in Fig.~\ref{fig1}(b). Here, $\bm{R}_0$ is a vector connecting
two neighboring mass points in equilibrium. In this paper, we adopt an
assumption that all springs have good linearity, i.e., the
energy of the spring $U_s$ is always written as
\begin{equation}
 U_s=\frac{1}{2}\kappa (l-l_0)^2,
\end{equation}
where $l$ is the length of the spring at that moment, independent of
the value of $l$. Then, up to the second order in
 $\delta{\bm{x}}\equiv\bm{x}'-\bm{x}''$, which is required to
 investigate harmonic oscillation around the equilibrium alignment, we
 have 
\begin{equation}
 U_s=\frac{1}{2}\kappa\bigl((R_0-l_0)^2+2(R_0-l_0)\hat{\bm{R}}_0\cdot\delta\bm{x}+\delta{x}_\mu\gamma^{\mu\nu}_{\hat{\bm{R}}_0}\delta{x}_\nu\bigr)
\end{equation}
with 
\begin{equation}
 \gamma^{\mu\nu}_{\bm{R}_0}=(1-\eta)\delta^{\mu\nu}
  +\eta \hat{R}^\mu_0\hat{R}^\nu_0,
\end{equation}
and $\hat{\bm{R}}_0=\bm{R}_0/|\bm{R}_0|$. Summation over indices
 appearing as a pair is assumed ($\mu,\nu=x,y$). This formula indicates that when
 $\eta=1$, $U_s$ strongly depends on the angle between $\bm{R}_0$ and
 $\delta{\bm{x}}$, while when $\eta=0$, $U_s$ is independent of
the direction of $\delta{\bm{x}}$. For $\eta=1$, no force is given by a
 spring if $\delta{\bm{x}}$ is normal to $\bm{R}_0$, since such a motion only
 leads to the rotation of the spring and does not increase the elastic
 energy in the lowest order approximation. On the other hand, for
 $\eta=0$, there is finite force even for
 $\delta{\bm{x}}=0$, and then the motion with
 $\delta{\bm{x}}\perp\bm{R}_0$ can increase the energy since there is a
 finite component of force normal to $\bm{R}_0$ for such motion. In short,
 the value of $\eta$ controls a ratio of strength of the restoring force
 for $\delta{\bm{x}}\parallel\bm{R}_0$ and $\delta{\bm{x}}\perp\bm{R}_0$. In
 other words, it controls the relation between the longitudinal and
 the transverse wave modes.

Then the Lagrangian of this system becomes
\begin{equation}
 \mathcal{L}=T-V_1-V_2-V_3,
\end{equation}
with
\begin{align}
 T&=\frac{1}{2}\sum_{\bm{R}a}\dot{x}^\mu_{\bm{R}a}\dot{x}^\mu_{\bm{R}a},\\
 V_1&=\frac{1}{2}\kappa^{0}\sum_{\bm{R}a}x^\mu_{\bm{R}a}x^\mu_{\bm{R}a},\\
 V_2&=\frac{1}{2}\kappa\sum_{\langle{\bm{R}'a\bm{R}b}\rangle}
 (x^\mu_{\bm{R}'a}-x^\mu_{\bm{R}b})
 \gamma^{\mu\nu}_{\tilde{\bm{R}}'_a-\tilde{\bm{R}}_b}
(x^\nu_{\bm{R}'a}-x^\nu_{\bm{R}b}),\\
 V_3&=\kappa\sum_{\langle{\bm{R}'a\bm{R}b}\rangle}\Delta^\mu_{\tilde{\bm{R}}'_a-\tilde{\bm{R}}_b}(x^\mu_{\bm{R}'a}-x^\mu_{\bm{R}b}),
\end{align}
where $\tilde{\bm{R}}_a$ is a position of the sublattice $a$ associated
 with the lattice point at $\bm{R}$, and
 $\Delta^\mu_{\bm{R}}=(R_0-l_0)R^\mu/|\bm{R}|$. Summation over
 $\langle{\bm{R}'a\bm{R}b}\rangle$ means that we pick up pairs of
the nearest neighbor mass points. 
The term $T$ represents a kinetic energy. The term $V_1$ is contributed
 from the potential energy other than the springs connecting the mass
 points, expanded in a series of
$\bm{x}_{\bm{R}a}$ up to the second order. For simplicity, we assume
 that this potential is isotropic and sublattice independent. The terms $V_2$ and $V_3$ come from
 the elastic energy of the springs. As far as the equilibrium is
 achieved for $\bm{x}_{\bm{R}a}=0$, the term $V_3$ is identically zero,
 and does not contribute to the equation of motion. 
 Assuming the periodic boundary condition and introducing
 $u^\mu_{\bm{k}a}$ as
\begin{equation}
 x^\mu_{\bm{R}a}=\frac{1}{N}\sum_{\bm{k}}\mathrm{e}^{\mathrm{i}\bm{k}\cdot\bm{R}}u^\mu_{\bm{k}a}, 
\end{equation}
the Lagrangian is rewritten as
\begin{align}
 \mathcal{L}&=\frac{1}{N}\sum_{\bm{k}}\mathcal{L}_{\bm{k}}\\
 \mathcal{L}_{\bm{k}},
 &=\frac{1}{2}\sum_a\dot{u}^\mu_{\bm{k}a}\dot{u}^\mu_{-\bm{k}a}
 -\frac{1}{2}\sum_{ab}\Gamma^{\mu\nu}_{ab}u^\mu_{\bm{k}a}u^\nu_{-\bm{k}b},
\end{align}
where $\Gamma^{\mu\nu}_{ab}=(\hat{\Gamma}(\bm{k}))_{a\mu;b\nu}$. Here, 
\begin{equation}
 \hat{\Gamma}(\bm{k})=\bigl(\kappa^{0}+3\kappa(1-\frac{\eta}{2})\bigr)\hat{1}
  +
  \begin{pmatrix}
   \hat{0}&\hat{\Gamma}_{AB}(\bm{k})\\
   \hat{\Gamma}_{AB}(-\bm{k})&\hat{0}\\
  \end{pmatrix},
\end{equation}
and 
\begin{equation}
 \hat{\Gamma}_{AB}(\bm{k})=-\kappa(\hat{\gamma}_3+\mathrm{e}^{-\mathrm{i}\bm{k}\cdot\bm{a}_1}\hat{\gamma}_1+\mathrm{e}^{-\mathrm{i}\bm{k}\cdot\bm{a}_2}\hat{\gamma}_2),
\end{equation}
with 
\begin{align}
 \hat{\gamma}_1&\equiv\hat{\gamma}_{\bm{R}_0^{(1)}}=
  (1-\eta)
  \begin{pmatrix}
   1&0\\
   0&1
  \end{pmatrix}
  +\eta
  \begin{pmatrix}
   \frac{3}{4}&\frac{\sqrt{3}}{4}\\
   \frac{\sqrt{3}}{4}&\frac{1}{4}
  \end{pmatrix},\\
 \hat{\gamma}_2&\equiv\hat{\gamma}_{\bm{R}_0^{(2)}}=
  (1-\eta)
  \begin{pmatrix}
   1&0\\
   0&1
  \end{pmatrix}
  +\eta
  \begin{pmatrix}
   \frac{3}{4}&-\frac{\sqrt{3}}{4}\\
   -\frac{\sqrt{3}}{4}&\frac{1}{4}
  \end{pmatrix},\\
 \hat{\gamma}_3&\equiv\hat{\gamma}_{\bm{R}_0^{(3)}}=
  (1-\eta)
  \begin{pmatrix}
   1&0\\
   0&1
  \end{pmatrix}
  +\eta
  \begin{pmatrix}
   0&0\\
   0&1
  \end{pmatrix}.
\end{align}
[See Fig.~\ref{fig1}(c) for the definitions of $\bm{R}_0^{(i)}$.]

Now equation to be solved becomes
\begin{equation}
  \ddot{u}^\mu_{\bm{k}a}+\sum_{b}\Gamma_{ab}^{\mu\nu}(\bm{k})u^\nu_{\bm{k}b}=0.
\end{equation}
We seek for the solution periodic in time by introducing a new
dynamical variable $\phi_{a\mu}(\bm{k})$ as
$u^\mu_{\bm{k}a}=\mathrm{e}^{\mathrm{i}\omega{t}}\phi_{a\mu}(\bm{k})$. Then,
what we have to solve is
\begin{equation}
 -\omega^2\phi_{a\mu}(\bm{k})
  +\sum_{b}\Gamma^{\mu\nu}_{ab}(\bm{k})\phi_{b\nu}(\bm{k})=0,
\end{equation}
and to have a nontrivial solution, we must have
\begin{equation}
 \text{Det}[-\omega^2\hat{1}+\hat{\Gamma}(\bm{k})]=0. \label{eq:det}
\end{equation}
Practically, this is done by diagonalizing $\hat{\Gamma}(\bm{k})$. In that
sense, it is possible to map the problem to a quantum mechanical one, by
regarding $\hat{\Gamma}(\bm{k})$ as a quantum Hamiltonian.

Here, we make a comment on the symmetry of
$\hat{\Gamma}(\bm{k})$. Except the constant diagonal elements,
$\hat{\Gamma}(\bm{k})$ has no term connecting a same kind of sublattice
components. Since the constant diagonal elements only add a constant
contribution to $\omega^2$, we say that $\hat{\Gamma}(\bm{k})$ has a
``chiral'' symmetry, since $\hat{\Gamma}(\bm{k})$ anticommute with
$\hat{\Upsilon}=\text{diag}(1,1,-1,-1)$ if the constant diagonal part is
subtracted. That is,
$\hat{\Gamma}'(\bm{k})=\hat{\Gamma}(\bm{k})-(\kappa^{0}+3\kappa(1-\frac{\eta}{2}))\hat{1}$
satisfies
\begin{equation}
 \{\hat{\Gamma}'(\bm{k}),\Upsilon\}=0.\label{chiral_symm_def}
\end{equation}
In fermionic systems, the chiral symmetry plays important roles in
various situations. For instance, it stabilizes Dirac cones in 2D cases
\cite{Hatsugai20091061}.

Now, the problem is quite similar to the phonon problem in
graphene
\cite{PhysRevB.65.155405,PhysRevB.67.035401,Wirtz2004141,PhysRevB.71.205214,PhysRevB.77.125401}. However,
what is essential in the following arguments is to
control $\eta$ in a wide range, which is difficult to realize in actual
graphene. Furthermore, the effective model for graphene phonon does not
respect the ``chiral symmetry'', which plays essential role in
considering the topological origin of the edge modes. This is because if
the vibrational motion in graphene is modeled using a spring-mass model,
it requires springs connecting the next nearest neighbor sites and
more. Lastly, our model neglects the motion in perpendicular to the
plane of the honeycomb lattice.

\begin{figure*}[tbp]
 \begin{center}
  \includegraphics{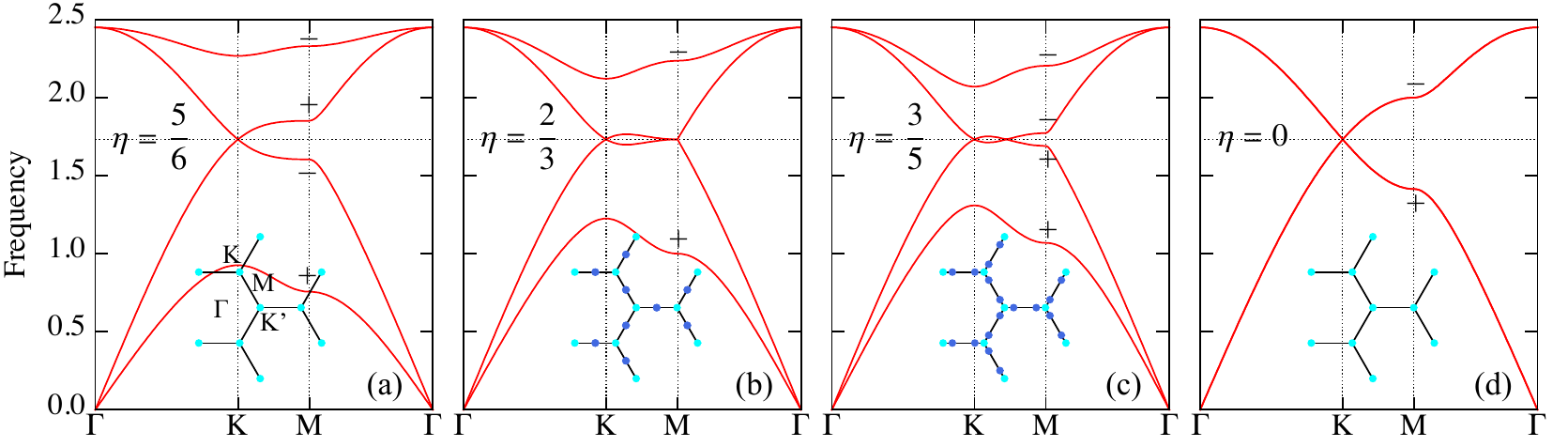}
  \caption{Dispersion relations for several $\eta$. (a) $\eta=5/6$. (b)
  $\eta=2/3$. (c) $\eta=3/5$. (d) $\eta=0$. Insets show the positions of
  the Dirac cones in the Brillouin zone. Cyan and blue dots represent
  the Dirac cones. $+$ and $-$ signs on the M-point are eigenvalues of the
  inversion operator Eq.~\eqref{inversion}. The spring constant $\kappa$
  is scaled as $\kappa=\kappa_0/(1-\eta/2)$.}\label{fig2}
 \end{center}
\end{figure*}
\section{Manipulation of Dirac Cones}
\subsection{Bulk Frequency Spectrum}
Now, let us investigate dispersion relation of $\omega$. Before going to
general values of $\eta$, we consider two limits $\eta=0$ and $\eta=1$. 
For $\eta=0$, $\hat{\Gamma}_{AB}(\bm{k})$ becomes
\begin{equation}
 \hat{\Gamma}_{AB}(\bm{k})=-\kappa(1+\mathrm{e}^{-\mathrm{i}\bm{k}\cdot\bm{a}_1}+\mathrm{e}^{-\mathrm{i}\bm{k}\cdot\bm{a}_2})
  \begin{pmatrix}
   1&0\\
   0&1
  \end{pmatrix}.
\end{equation}
Then $\omega^2$ looks exactly like a dispersion relation of the nearest
neighbor tight-binding model for graphene (NN-graphene) but with extra
double degeneracy. In $\eta=0$ limit, there is no distinction between
the longitudinal and transverse wave mode, and in this 
case, oscillation in $x$- and $y$-direction decouple, which leads
to doubled NN-graphene dispersion. On the other hand for $\eta=1$,
$\hat{\Gamma}_{AB}(\bm{k})$ becomes
\begin{equation}
 \hat{\Gamma}_{AB}(\bm{k})=-\kappa
  \begin{pmatrix}
   \frac{3}{4}(\mathrm{e}^{-\mathrm{i}k_1}+\mathrm{e}^{-\mathrm{i}k_2})&\frac{\sqrt{3}}{4}(\mathrm{e}^{-\mathrm{i}k_1}-\mathrm{e}^{-\mathrm{i}k_2})\\
   \frac{\sqrt{3}}{4}(\mathrm{e}^{-\mathrm{i}k_1}-\mathrm{e}^{-\mathrm{i}k_2})&\frac{1}{4}(\mathrm{e}^{-\mathrm{i}k_1}+\mathrm{e}^{-\mathrm{i}k_2})+1
  \end{pmatrix}
\end{equation}
with $k_i=\bm{k}\cdot\bm{a}_i$. Then, as shown in
Ref.~\onlinecite{Coriolis:2014}, the second and third bands of $\omega^2$, 
dispersion is same as NN-graphene, but the first and the fourth bands
are exactly flat and stick to the top and bottom of the other
bands. Interestingly, $\hat{\Gamma}(\bm{k})$ at $\eta=1$ is essentially
same as the Hamiltonian of $p$-orbital honeycomb optical lattice model
introduced in Ref.~\onlinecite{PhysRevA.83.023615}. 

Now, we follow the evolution of the dispersion relation as $\eta$
changes from 1 to 0. In the following, we always scale the spring
constant $\kappa$ as $\kappa=\kappa_0/(1-\eta/2)$ with $\kappa_0=1$, 
in order to make the total band width of the frequency dispersion
constant. This scaling also
makes the frequency of the Dirac point constant. The dispersion
relation for $\eta=5/6$ is shown in
Fig.~\ref{fig2}(a). It is similar to the one for $\eta=1$ and the Dirac cone is still
there at the K-point (and K'-point), but the first and the fourth bands
gains finite dispersion and they are no longer flat. At $\eta=2/3$, as
shown in Fig.~\ref{fig2}(b), the gap at the M-point in between the
second and the
third bands closes. In the vicinity of the M-point, the dispersion is
linear in $\Gamma$-M direction but quadratic in M-K direction. Such a
linear vs quadratic dispersion relation is often characteristic for
merging of a pair of Dirac cones
\cite{springerlink:10.1140/epjb/e2009-00383-0}. In fact, soon after
$\eta$ becomes
less than 2/3, there appear two new Dirac cones, other than those at the
K- and K'- points, on M-K lines. [See Fig.~\ref{fig2}(c).] Due to the
three fold rotational symmetry, there appear six new Dirac cones in the whole
Brillouin zone in total. The new Dirac cones move from the M-point to
the K-point as $\eta$ approaches to zero. For $\eta<2/3$,
especially for small $\eta$ where the positions of the new Dirac cones
get close to the K- and K'-points, the dispersion relation looks like
the one for bilayer graphene with the trigonal warping
\cite{PhysRevLett.96.086805}. Finally, at
$\eta=0$, the new Dirac cones are absorbed to the originally existing
Dirac cones at the K- and K'-points, and the dispersion becomes
degenerated NN-graphene dispersion. 

The transition at $\eta=2/3$ that is associated with the Dirac cone
merging can be characterized by the ``Herring number''. For
electronic systems, the Herring number is defined as
\cite{PhysRev.52.365}
\begin{equation}
 N_H=\frac{1}{2}\sum_{r=1}^{2^d}[N^+(\bm{k}_r)-N^-(\bm{k}_r)],
\end{equation}
where $d$ is spatial dimension, $\bm{k}_r$s are the time reversal
invariant momenta, and $N^{\pm}(\bm{k}_r)$ is the number of occupied
states with parity eigenvalue $\pm 1$ at $\bm{k}_r$. This number is also
used by Fu and Kane to obtain $Z_2$ topological number of the inversion
symmetric topological insulator with the spin--orbit coupling, or to
check the existence of Dirac cones without the spin--orbit
coupling \cite{PhysRevB.76.045302}. For 2D cases, parity of $N_H$
determines the parity of the number of Dirac
cones in half of the Brillouin zone as explained below. In
Ref.~\onlinecite{PhysRevB.88.245126}, in order to detect Dirac cones in
generic 2D systems, we have used the Berry phase defined as
\begin{equation}
 \mathrm{i}\theta(k_{\parallel})=\sum_{n\in\text{filled}}
  \int_{-\pi}^{\pi}\mathrm{d}k_{\perp}
  \langle u_{nk_\parallel k_\perp}|\nabla_{k_\perp}
  |u_{nk_\parallel k_\perp}\rangle,
  \label{Berry_def}
\end{equation}
where $k_\parallel$ and $k_\perp$ are two momenta in independent
 directions, and $|u_{nk_\parallel k_\perp}\rangle$ is a Bloch wave
 function \cite{PhysRevB.84.195452,PhysRevB.88.245126}. When the system
 respects both of the time reversal and
 spatial inversion symmetry, $\theta(k_\parallel)$ is quantized into $0$
 or $\pi$ for the spinless case \cite{PhysRevB.88.245126}. As far as the
 quantization is kept,
 $\theta(k_\parallel)$ has to show a jump when it changes as a function
 of $k_\parallel$, but the jump should be associated with a
 singularity of the band structure that is nothing more than a Dirac
 cone. Having this in mind, we can say that if
 $\theta(0)\neq\theta(\pi)\text{ mod }2\pi$, odd number of
 Dirac cones exist in the region $0<k_\parallel<\pi$, i.e., half of
 the Brillouin zone, while if $\theta(0)=\theta(\pi)\text{ mod }2\pi$,
 the number of Dirac cones in the half Brillouin zone is even. On the other hand, the inversion symmetry gives us
 a relation \cite{PhysRevB.76.045302,doi:10.7566/JPSJ.82.034712}
 \begin{equation}
  \mathrm{e}^{\mathrm{i}(\theta(\pi)-\theta(0))}=\prod_{r=1}^4(-1)^{N^-(\bm{k}_r)}=(-1)^{N_H}.
   \label{parity_formula_2}
 \end{equation}
Therefore, the Herring number has an ability to detect the number of Dirac
cones. This idea has been applied to the 2D organic Dirac fermion
systems \cite{doi:10.7566/JPSJ.82.033703,doi:10.7566/JPSJ.82.034712}.
Now, we are handling a classical mechanical system, not an electronic
system, but the Herring number is still useful. We have seen that the
number
of Dirac points in the half Brillouin zone is odd (one) in $\eta>2/3$
and even (four) in $\eta<2/3$. This
difference should be captured by the Herring number. In
Figs.~\ref{fig2}(a)-\ref{fig2}(d), the eigenvalues of the inversion
operator $\hat{U}_I$, which is defined as 
\begin{equation}
 \hat{U}_I=
  \begin{pmatrix}
    0&0&1&0\\
    0&0&0&1\\
    1&0&0&0\\
    0&1&0&0
  \end{pmatrix}\label{inversion}
\end{equation}
at the M-point are indicated for each band. We notice
that at the transition, $\eta=2/3$, the positive and negative
eigenvalues at the M-point are interchanged. Since there are three
distinct M-points in the Brillouin zone, this interchange results in
the parity change of the Herring number.
Note that Herring originally considers only three-dimensional cases, and in his paper,
$N_H$ is derived to detect the number of degeneracy loops in 3D
Brillouin zone, instead of Dirac cones.

\begin{figure}[tbp]
 \begin{center}
  \includegraphics{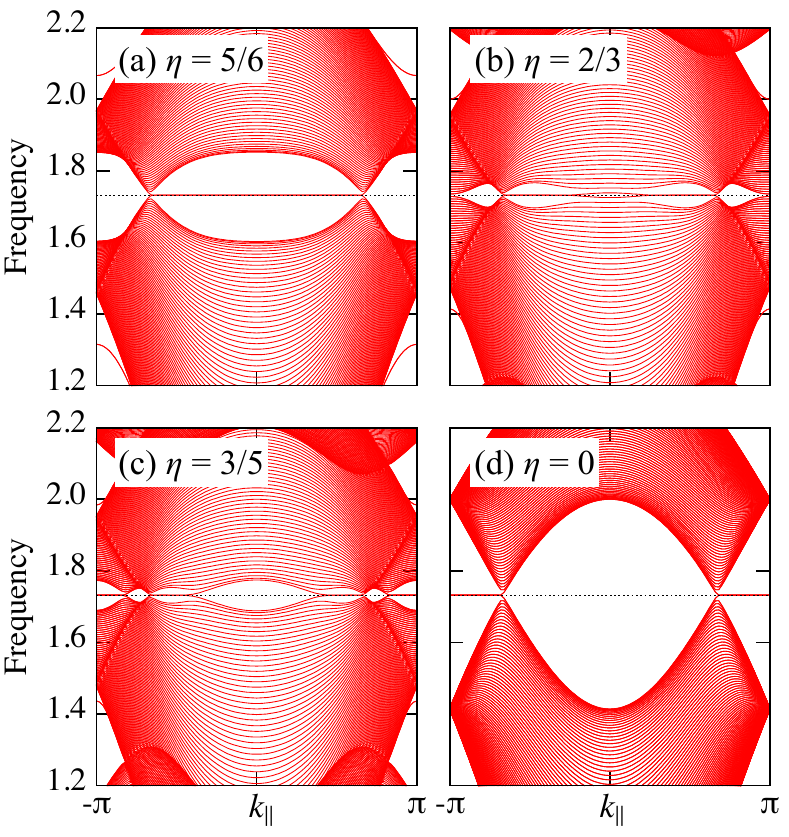}
  \caption{Edge spectra as a function of the momentum along the edge for
  several $\eta=1$ with the fixed boundary condition. (a) $\eta=5/6$.
  (b) $\eta=2/3$. (c) $\eta=3/5$. (d) $\eta=0$.}\label{fig3}
 \end{center}
\end{figure}
\subsection{Edge States and Symmetry Protection}
\begin{figure}[tbp]
 \begin{center}
  \includegraphics{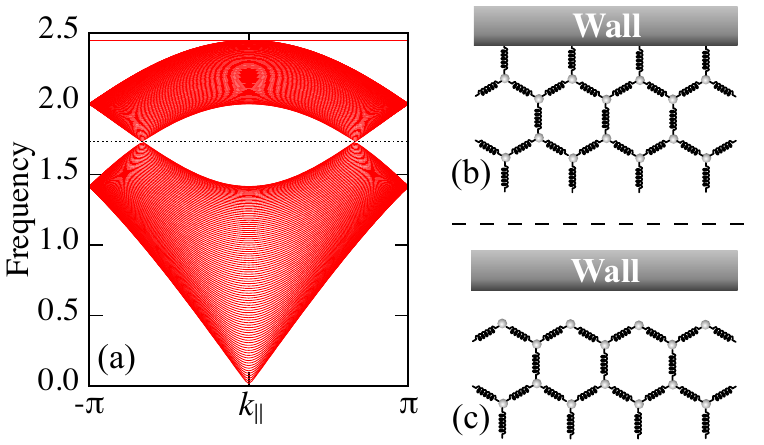}
  \caption{(a) Edge spectrum for the free boundary condition with
  $\eta=1$. (b,c) Schematic pictures of the (b) fixed and (c) free
  boundary conditions.} \label{fig4}
 \end{center}
\end{figure}

Next, we investigate edge states. In this case, we only apply the
Fourier transformation in one direction and for the other direction,
real space representation is used. Then, we obtain the equation similar
to Eq.~\eqref{eq:det}, but with larger number of components in the
matrices and vectors. In the following, we concentrate on the zigzag edge. In
addition, we use the fixed boundary condition, that is, the mass points
at the boundary are connected to the springs that are fixed to the wall,
instead of simply removing the springs at the boundary. [See
Figs.~\ref{fig4}(b) and \ref{fig4}(c).] Importance of
the boundary condition is discussed later. 
The obtained frequency spectra for several values of $\eta$ as functions
of $k_\parallel$, momentum parallel to the edge, are shown in
Figs.~\ref{fig3}(a)-\ref{fig3}(d). 
For $\eta=5/6$, as in
the case of $\eta=1$\cite{Coriolis:2014}, we observe in-gap edge modes
near $k_\parallel=0$. The edge modes forms a flat band, since the
``chiral symmetry'' is preserved in this case. At $\eta=0$, the system
exactly inherits the property of NN-graphene. For instance, the edge
modes are found near $k_\parallel=\pi$ instead of $k_\parallel=0$ found
for $\eta>2/3$. In short, the position of the edge modes in the edge 
Brillouin zone is switched from near $k_\parallel=0$ to
$k_\parallel=\pi$ as $\eta$ changes from $1$ to $0$. We observe that the
new Dirac cones emerged from the M-points bring new edge modes near
$k_\parallel=\pi$ and take away the originally existing edge modes near
$k_\parallel=0$. Again, note that our model does not necessary leads to
the results exactly same as the case of the actual graphene phonon
problems \cite{PhysRevB.81.165418,PhysRevB.81.174117}.

\begin{figure*}[tbp]
 \begin{center}
  \includegraphics{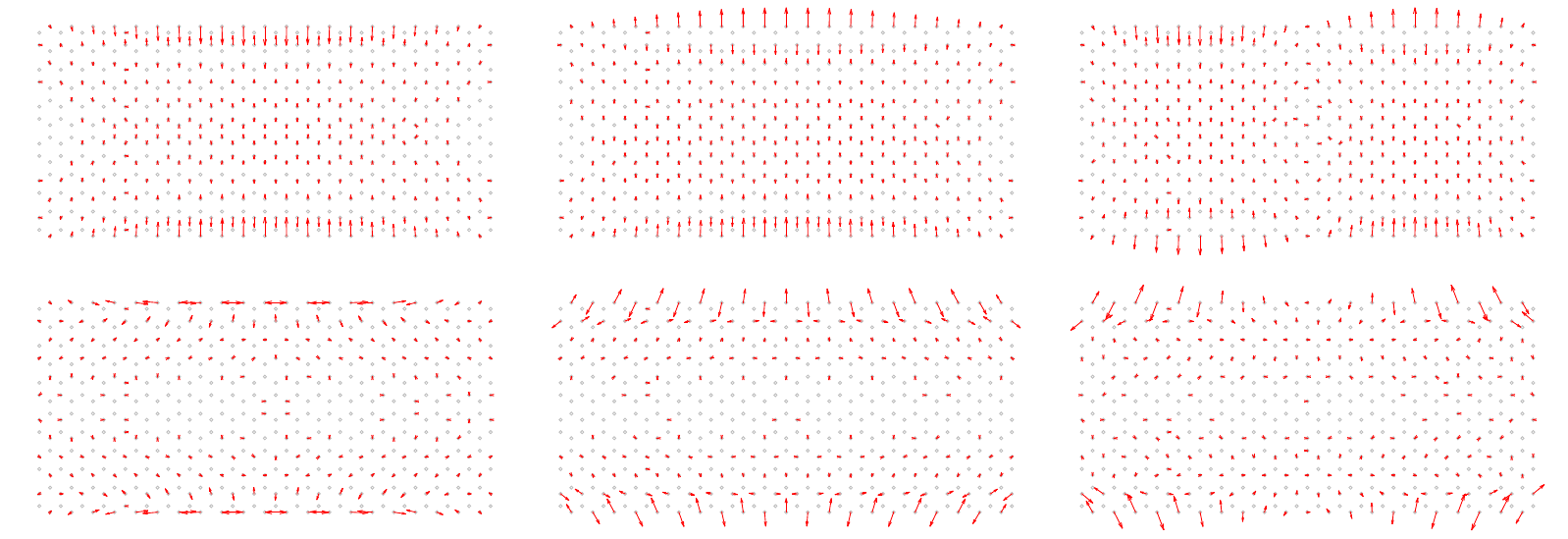}
  \caption{Real space picture of the typical eigenmodes that are
  localized to the edge. Upper panels: $\eta=1$. Lower panels:
  $\eta=1/3$. For $\eta=1$ ($\eta=1/3$), the neighboring mass points at
  the edge oscillate in-phase (anti-phase).}\label{fig5}
 \end{center}
\end{figure*}
To see the importance of the boundary condition, $\omega$ for $\eta=1$
obtained with the free boundary condition, where the springs at the
boundary are simply removed, is shown in Fig.~\ref{fig4}. Note that
$\eta\neq 1$
does not go well with the free boundary condition since the finite
tension results in the deformation of the equilibrium lattice as the
springs at the boundary becomes absent. Then, we have to take
the deformation into account, but instead of doing that, we focus on
$\eta=1$. Figure~\ref{fig4} shows that there is no edge mode
found as in-gap states. This difference between the two boundary
conditions is deeply related to symmetry protection of topological
phases, which is one of the most important concept in modern study of
topological phenomena. In the real graphene case, it is possible to
think a topological origin of the edge modes
\cite{PhysRevLett.89.077002}. For that, the chiral symmetry plays
an important role in defining the bulk topological number, the quantized
Berry phase, and protecting in-gap edge states. For the fixed boundary
condition, the ``chiral symmetry'', Eq.~\eqref{chiral_symm_def}, is
preserved even after the boundary
is introduced, since the springs connected to the mass points at the
boundary is left. On the other hand, the free boundary condition does
not respects the ``chiral symmetry'' since the absence of the springs at
the boundary leads to the {\it nonuniform} diagonal elements in
$\hat{\Gamma}(k)$. This indicates that even the bulk topological
property is same, the topological edge states are not necessarily
preserved for the free boundary condition, in contrast to the fixed
boundary condition, which demonstrates the  manifestation of the
symmetry protection. 

Now, we know that the edge modes for $\eta\sim 1$ and $\eta\sim 0$ has
different momenta along the edge. Then, the oscillation pattern in the
real space should look different. In order to convince this expectation,
we perform a calculation using a rectangular system with long zigzag
edges and short armchair edges. This calculation also gives a result to
the situation that is easily accessed in experiment. In Fig.~\ref{fig5},
the typical eigenmodes whose frequency is close to $\sqrt{3}$, a
frequency of the bulk Dirac points, is shown for $\eta=1/3$ and $\eta=1$.
For $\eta=1$, the mass points at the boundary
basically oscillate in phase with the neighboring mass points on the
edge with large wave length background, reflecting
the fact that the edge modes are existing near $k_\parallel=0$ in the
momentum space. On the other hand, for $\eta=1/3$, the neighboring mass
points on the edge tend to oscillate anti-phase, reflecting the edge
mode position in the momentum space. For $\eta=1$, the mass points at
the boundary oscillate approximately normal to the boundary, while for
$\eta=1/3$, oscillation in parallel to the boundary is also allowed. In
Fig.~\ref{fig5}, only the typical eigenmodes are shown, but we
have confirmed that the above statements are basically valid for the
mode near $\omega=\sqrt{3}$. Thus, the oscillation pattern of the
eigenmode localized at the boundary can be used to detect the phase
transition associated with the Dirac cone merging. This is important
since the observation of the edge mode with finite size system may be
much easier than measuring the dispersion relation itself in detail. 

\section{Chern Number and Chiral Edge Modes}
Lastly, we consider effects of rotation, which induces time reversal
symmetry breaking. Uniform rotation as a whole brings two new terms in
the Lagrangian, one is the centrifugal force and the other is the
Coriolis force. For simplicity, the centrifugal force is neglected. The
centrifugal force is second order in the angular frequency of the
rotation $\Omega$, while the Coriolis force is first order in $\Omega$, 
and the second order term can be neglected in the situation that
$\Omega$ is regarded as a small quantity. Then, taking account of the
Coriolis force only, the equation to be solved becomes \cite{Coriolis:2014}
\begin{equation}
 \mathrm{Det}[-\omega^2\hat{1}+2\mathrm{i}\omega\hat{\Omega}+\hat{\Gamma}(\bm{k})]=0, \label{eq:detC}
\end{equation}
where
\begin{equation}
 \hat{\Omega}=
  \begin{pmatrix}
   \hat{\Omega}_0&\hat{0}\\
   \hat{0}&\hat{\Omega}_0\\
  \end{pmatrix},\quad
  \hat{\Omega}_0=
  \begin{pmatrix}
   0&\Omega\\
   -\Omega&0
  \end{pmatrix}.
\end{equation}
In this case it is no longer possible to obtain allowed $\omega$ by
diagonalizing $\hat{\Gamma}(\bm{k})$. Instead, it is
possible to obtain allowed $\omega$ by explicitly evaluating the
determinant analytically, and solving a quartic equation of
$\omega^2$. When $\kappa^{0}$ is large, the equation to be
solved approximately becomes
\begin{equation}
 \mathrm{Det}[-\omega^2\hat{1}+2\mathrm{i}\sqrt{\kappa^{0}}\hat{\Omega}+\hat{\Gamma}(\bm{k})]=0, 
\end{equation}
and diagonalization of
$2\mathrm{i}\sqrt{\kappa^{0}}\hat{\Omega}+\hat{\Gamma}(\bm{k})$ is 
sufficient to obtain $\omega$. Interestingly,
$2\mathrm{i}\sqrt{\kappa^{0}}\hat{\Omega}+\hat{\Gamma}(\bm{k})$ is exactly
like the Hamiltonian for $p$-orbital honeycomb optical lattice in
Ref.~\onlinecite{PhysRevA.83.023615}. 

In Ref.~\onlinecite{Coriolis:2014}, it is shown that the finite $\Omega$
induces a gap at the Dirac point for $\eta=1$. By solving
Eq.~\eqref{eq:detC}, it is
confirmed that such a gap opening remains at work as $\eta$ is gradually
reduced from 1. At some critical point $\eta=\eta_c$, the gap is closed
at the M-point. Then, for $\eta$ less than $\eta_c$, the system is again
fully gapped except $\eta=0$ where the gap is again absent. $\eta_c$
depends on $\Omega$, but it is close to 2/3 in the small $\Omega$ limit.

\begin{figure}[tbp]
 \begin{center}
  \includegraphics{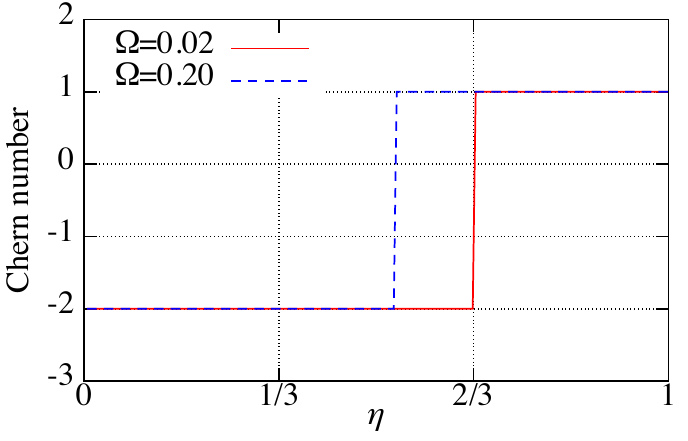}
  \caption{The Chern number as a function of $\eta$ for $\Omega=0.02$
  and 0.20.}\label{fig6}
 \end{center}
\end{figure}

After obtaining $\omega$, eigenmodes are derived as nontrivial solutions
for
\begin{equation}
 [-\omega^2\hat{1}+2\mathrm{i}\omega\hat{\Omega}+\hat{\Gamma}(\bm{k})]\bm{\phi}_{n\bm{k}}=0
\end{equation}
Then, regarding the solution $\bm{\phi}_{n\bm{k}}$ as a Bloch wave
function $|u_{n\bm{k}}\rangle$, we can define the Chern number $C$ as
\begin{equation}
 C=\sum_{n=1,2}\frac{1}{2\pi}\iint\mathrm{d}k_1\mathrm{d}k_2
  \biggl(
  \frac{\partial \bm{\phi}^\dagger_{n\bm{k}}}{\partial k_1}
  \cdot\frac{\partial \bm{\phi}_{n\bm{k}}}{\partial k_2}
  -
  \frac{\partial \bm{\phi}^\dagger_{n\bm{k}}}{\partial k_2}
  \cdot\frac{\partial \bm{\phi}_{n\bm{k}}}{\partial k_1}
  \biggr)
\end{equation}
except at $\eta=0$ and $\eta=\eta_c$ where the gap closing occurs. The
summation is taken over $n=1,2$, since we are now focusing on the gap
between the second and third bands. The
Chern number can be used as a topological order parameter for this
system. There is a well established method to evaluate this quantity
numerically \cite{doi:10.1143/JPSJ.74.1674}. In Fig.~\ref{fig6}, $\eta$ dependence of the numerically obtained
Chern number for $\Omega=0.02$ and 0.20 is shown. The result at $\eta=1$
is consistent with the previous work \cite{Coriolis:2014}. For small
$\Omega$, the transition from Chern number 1 to $-2$ is observed near
$\eta=2/3$. This is well understood from the fact that each Dirac point
carries one-half contribution to the Chern number. For $\Omega=0.2$,
the transition point deviates from 2/3, but the transition from 1 to $-2$
still exists. Transition involving change of the Chern number is
reported in Ref.~\onlinecite{Coriolis:2014}, but here we propose a
transition with completely different mechanism. 

In order to observe physical phenomena associated with this transition
of the Chern number, we perform a calculation with finite system with
a triangular shape having zigzag edges. The Chern number for the
mechanical system itself is not a physical observable but the
corresponding edge states reflecting nontrivial bulk can be observed
experimentally \cite{PhysRevLett.71.3697}. Here, again, the fixed boundary
condition is applied, i.e., the system must be fit in a triangular frame
supporting mass points at the boundary by springs. We assume that the
system is at rest, or no vibration mode is excited at $t=0$. Then, we
pick up one of the mass points and apply force that is sinusoidal in
time with angular frequency $\Omega'$, i.e., we consider a forced
oscillation problem. Now, equation to be solved can be rewritten as
\begin{equation}
 \frac{\mathrm{d}\bm{X}(t)}{\mathrm{d}t}=\hat{A}\bm{X}(t)+\bm{f}(t)
\end{equation}
where $\bm{X}(t)={}^t(\dot{\bm{x}}(t),\bm{x}(t))$ and
\begin{equation}
 \hat{A}=
  \begin{pmatrix}
   2\hat{\Omega}&-\hat{\Gamma}\\
   \hat{1}& 0
  \end{pmatrix}.
\end{equation}
Here, $\hat{\Gamma}$ is a dynamical matrix and $\hat{\Omega}$ is defined
as 
\begin{equation}
 \hat{\Omega}=
  \begin{pmatrix}
   \hat{\Omega}_0&\hat{0}&\cdots\\
   \hat{0}&\hat{\Omega}_0&\cdots\\
   \vdots&\vdots & \ddots
  \end{pmatrix}.
\end{equation}
$\bm{f}(t)$ denotes an external force, and is explicitly written as
$\bm{f}(t)=\bm{f}_0\sin\Omega't$. The elements of $\bm{f}_0$ have a
finite value when it corresponds to the picked mass point. A formal
solution of this equation is
\begin{equation}
 \bm{X}(t)=\int^t_0\mathrm{d}\tau
  \exp\bigl(\hat{A}(t-\tau)\bigr)\bm{f}(\tau) \label{formal_solution}
\end{equation}
when the system is at rest for $t=0$. Then, $\bm{X}(t)$ at any time is
evaluated by diagonalizing $\hat{A}$ and perform the integration in
Eq.~\eqref{formal_solution}. 
\begin{figure}[tbp]
 \begin{center}
  \includegraphics{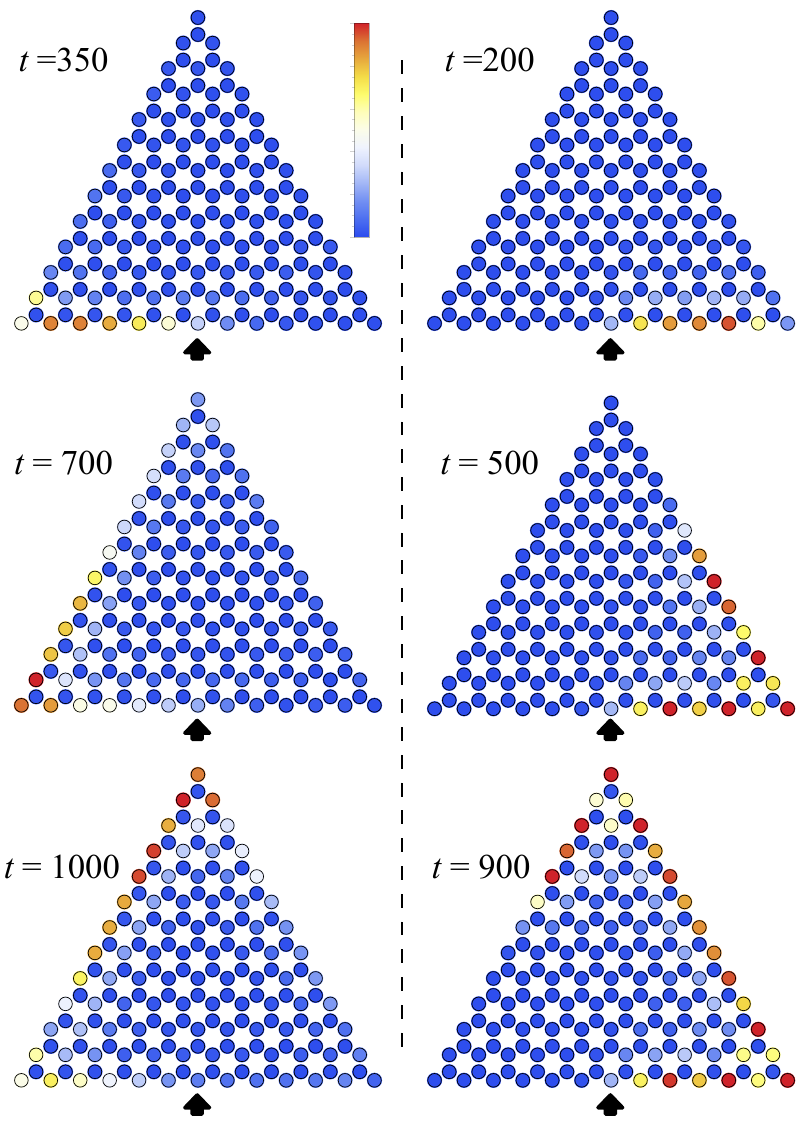}
  \caption{Snapshots of the time evolution of the system. The color map
  indicates the kinetic energy of each mass point averaged over the time
  range $2\pi/\sqrt{3}$. The external force is
  applied to the mass points indicated by thick arrows. Left panels:
  $\eta=1/3$ and $C=-2$. We observe clockwise motion of the edge
  modes. Right panels: $\eta=1$ and $C=1$. We observe counterclockwise
  motion of the edge modes.}\label{fig7}
 \end{center}
\end{figure}

The selected snapshots of the time
evolution for $\eta=1/3$ (The Chern number is $-2$) and $\eta=1$ (The
Chern number is $1$) calculated with $\Omega=0.05$ are shown in
Fig.~\ref{fig7}, where the color map
indicates the kinetic energy of each mass point averaged over time
period $2\pi/\Omega'_0$ with $\Omega'_0=\sqrt{3}\sim\Omega'$. In the
figure, thick arrows indicates the mass
point on which the external force is applied. Here, the direction of
force is chosen to be normal to the edge. We use $\Omega'=1.765$ for
$\eta=1/3$ and $\Omega'=1.732$ for $\eta=1$ in order to excite edge modes
efficiently. Owing to these choices, for both of $\eta=1/3$ and
$\eta=1$, the oscillation amplitude is localized near the edge. As in
the case of the rectangular system, the neighboring mass points at the
edge oscillate anti-phase for $\eta=1/3$ and in-phase for
$\eta=1$. Following the time evolution of the oscillation, we notice
that the oscillation amplitude propagates in the fixed direction,
which indicates existence of the chiral edge modes. When the ``wave
front'' of the oscillation reaches to the corner, it goes to the next
edge instead of being reflected. For $\eta=1/3$ with
the Chern number $-2$, the wave front shows clockwise motion, while for
$\eta=1$ with the Chern number $1$, it shows counterclockwise motion. An
interesting point of this
observation is that the difference between $\eta=1/3$ and $\eta=1$
resides in the relation between the transverse and longitudinal modes,
and the symmetry of the system is same for $\eta=1/3$ and
$\eta=1$. Nevertheless, it reverses the direction of the motion of the
wave front. 

Before closing, we make a comment on the choice of $\Omega'$. In
principles, the chiral edge mode is a gapless excitation. However, as we
are handling the finite size system, the gap is inevitably open, and we
have to be careful on choosing $\Omega'$ to excite the edge modes. There
are two kinds of finite size effects leading to gap opening. Firstly, the
finite size effect simply makes the energy levels discretized, but this
does not make much problems in exciting the edge modes. Secondly, in
some case, the edge modes localized on different edges are mixed by the
finite size effect and lead to a gap that is larger than the one
expected from the simple level discretization. In such a case, we have
to avoid to place $\Omega'$ in such a large gap in order to excite the
chiral edge modes effectively. For $\eta=1/3$, we encountered with the
situation that such a gap is induced, and then, careful choice of 
$\Omega'$ is essential to observe the behavior like Fig.~\ref{fig7}. 

\section{Summary}
To summarize, it is shown that the dispersion relation of the mechanical
graphene is dramatically changed by controlling only one parameter,
the spring tension at equilibrium. When there is no tension at
equilibrium, the frequency dispersion is characterized by Dirac cones at
the K- and K'-points and flat bands at the bottom and top of the
dispersion. As the tension at equilibrium is strengthen, extra Dirac
cones are created at the M-points. Then, the newly created Dirac cones
migrate from the M-points to the K- or K'-points, and the dispersion
becomes similar to the electronic dispersion of the bilayer graphene
with the trigonal warping. Eventually, the extra Dirac cones are
absorbed to the Dirac cones at the K- and K'-points, and the dispersion is
characterized by doubly degenerated Dirac cones in the strong tension
limit. The singular bulk
dispersion is reflected in the edge spectrum, which is confirmed by a
calculation with the ribbon geometry and the fixed boundary
condition. Corresponding to the generation and merging of the bulk Dirac
cones, the position of the edge states in the momentum space changes,
and this change can be detected by observing the real space pattern of
the oscillation of the edge modes. On the other hand, the free boundary
condition gives no edge modes. In the topological viewpoint, absence of
the edge mode for the free boundary condition is explained by breaking
of the ``chiral symmetry'' at the edge. These observation forms a good
example of the bulk--edge correspondence. Lastly, Coriolis force is
introduced to discuss the time reversal symmetry breaking. Then,
generation and merging of the Dirac cones found without Coriolis force
appear as a jump in the Chern number, which is defined regarding the
vector of normal modes as Bloch wave functions. Furthermore, it is
demonstrated that when the sign of the Chern number is flipped, then the
propagation direction of the chiral edge modes is really
flipped. Importantly, flip in the propagation direction is caused by
only controlling the tension at equilibrium, which keeps the symmetry of
the system intact. That is, the system looks same, but still the edge
modes propagate in the opposite directions. 

\section{Acknowledgments}
The work is partly supported by Grants-in-Aid for Scientific Research
No. 26247064 from JSPS and No. 25107005 from MEXT.

\end{document}